\documentclass[preprint,3p,times, twocolumn]{elsarticle}

\usepackage{graphicx}

\journal{Physics Letters B}

\begin{document}

\begin{frontmatter}

\title{Determination of hexadecapole ($\beta_{4}$) deformation of the light-mass nucleus $^{24}$Mg using quasi-elastic scattering measurements}

\author[BARC,HBNI,ND]{Y. K. ~Gupta\corref{cor1}}
\cortext[cor1]{Corresponding author, Email:ykgupta@barc.gov.in}

\author[BARC,HBNI]{B. K. Nayak}

\author[ND]{ U. Garg}

\author[KYOTO]{K. Hagino}

\author[ND]{ K. B. Howard}
\author[ND]{ N. Sensharma}
\author[Ankara,ND]{ M. {\c S}enyi{\u g}it}
\author[ND]{ W. P. Tan}
\author[ND]{ P. D. O'Malley}
\author[ND]{ M. Smith}

\author[BARC]{Ramandeep Gandhi}

\author[ND]{T. Anderson}
\author[ND]{R. J. deBoer}
\author[ND]{ B. Frentz}
\author[ND]{ A. Gyurjinyan}
\author[ND]{ O. Hall}
\author[ND]{ M. R. Hall}
\author[ND]{ J. Hu}
\author[ND]{ E. Lamere}
\author[ND]{ Q. Liu}
\author[ND]{ A. Long}
\author[ND]{ W. Lu}
\author[ND]{ S. Lyons}
\author[ND]{ K. Ostdiek}
\author[ND]{ C. Seymour}
\author[ND]{ M. Skulski}
\author[ND]{ B. Vande Kolk}

\address[BARC]{Nuclear Physics Division, Bhabha Atomic Research Centre, Mumbai - 400085, INDIA}
\address[HBNI]{Homi Bhabha National Institute, Anushaktinagar, Mumbai 400094, India}
\address[ND]{Physics Department, University of Notre Dame, Notre Dame, IN 46556, USA}
\address[KYOTO]{Department of Physics, Kyoto University, Kyoto 606-8502, Japan}
\address[Ankara]{Department of Physics, Faculty of Science, Ankara University, TR-06100 Tandogan, Ankara, Turkey}

\date{\today}

\begin{abstract}
Quasi-elastic scattering measurements have been performed using $^{16}$O and $^{24}$Mg projectiles off $^{90}$Zr at energies around the Coulomb barrier. Experimental data have been analyzed in the framework of coupled channels (CC) calculations using the code CCFULL. The quasi-elastic scattering excitation function and derived barrier distribution for $^{16}$O + $^{90}$Zr reaction are well reproduced by the CC calculations using the vibrational coupling strengths for $^{90}$Zr reported in the literature. Using  these vibrational coupling strengths, a Bayesian analysis is carried out for $^{24}$Mg + $^{90}$Zr reaction. The $\beta_{2}$ and $\beta_{4}$ values for $^{24}$Mg are determined to be $+0.43 \pm 0.02$ and $ - 0.11 \pm 0.02$, respectively. The $\beta_{2}$ parameter determined in the present work is in good agreement with results obtained using inelastic scattering probes. The hexadecapole deformation of $^{24}$Mg has been measured very precisely for the first time. Present results establish that quasi-elastic scattering could provide a useful probe to determine the ground state deformation of atomic nuclei.
\end{abstract}

\end{frontmatter}



Determining the ground state properties of atomic nuclei away from the $\beta$-stability line, such as the mass, spin, shape, half-life, electromagnetic moments, and many others, is among the primary foci of current nuclear physics research. Obtaining experimental values for these nuclear properties is very crucial for the benchmarking of macroscopic-microscopic and microscopic theories which guide the exploration of the nuclear chart. In this context, state-of-the-art Radioactive Ion Beam (RIB) facilities are being developed at premier labs across the globe. The primary bottleneck while studying the properties of the exotic nuclei is the low intensity of the RIBs in contrast to the stable beams.

Among many other properties, knowing precise information about the ground state deformation of the atomic nuclei is of fundamental importance. It is not only for their roles in heavy-ion reaction dynamics, but also to understand the microscopic interaction responsible for nuclear structure. In this context, static ground state deformations of atomic nuclei such as quadrupole ($\beta_{2}$), octopole ($\beta_{3}$), and hexadecapole  ($\beta_{4}$), are of vital significance.  Previously, electron-scattering \cite{Cooper1976,Horikawa1971, Akira1972}, Coulomb excitation \cite{FEWELL1979, Wollersheim1975}, proton-scattering \cite{Fabrici1980, Ichihara1987, swiniarski1969, Leo1979}, neutron-scattering \cite{Haouat1984}, deuteron-scattering \cite{TJIN1968, KISS1969}, $^{3}$He-scattering \cite{Kemper1971, Peterson1978}, $\alpha $-scattering \cite{Rebel1972, Harakeh1979}, heavy-ions \cite{Wilczynska1974,  Mittig1974}, and muonic X rays \cite{Powers1975} have been used  to determine the deformation of atomic nuclei experimentally. In comparison to lower order multipoles - quadrupole and octopole - the hexadecapole deformation is difficult to determine experimentally with a good precision, primarily because of its small magnitude.

Systematic studies of heavy-ion reaction dynamics have revealed that there is a strong interplay between nuclear structure effects and the relative motion of the two colliding nuclei \cite{Dasgupta, HaginoPTP2012}. In particular, during heavy-ion fusion, the  coupling of internal degrees of freedom of the fusing nuclei, such as vibrational (spherical), rotational (deformed), and particle transfer, gives rise to a distribution of fusion barriers instead of a single barrier \cite{Dasgupta, HaginoPTP2012}. These barrier distributions provide a fingerprint of nuclear structure effects of the colliding nuclei. For instance, a comparison of the fusion barrier distributions between the  $^{16}$O + $^{154}$Sm and the $^{16}$O + $^{186}$W systems has shown that a barrier distribution is sensitive to the sign of hexadecapole deformation parameter \cite{Leigh1995}. 

It has also been established that a representation of fusion-barrier distribution can be extracted from quasi-elastic (QEL) scattering, measured at backward angles \cite{Timmers1995, HaginoPRC2004}. If colliding partners are chosen appropriately such that the transfer channel coupling strength is weak, QEL scattering can be used as a probe to determine quantitatively the strengths of collective degrees of freedom.  However, its applicability to determine the ground state deformation has been demonstrated only in a few cases so far; only in the heavy-mass region of rare earths \cite{Jia2014}.

In the light mass region of the 2s-1d shell, the quadrupole deformation ($\beta_{2}$) has been determined to a good precision using the aforementioned inelastic-scattering probes. In particular, $\beta_{2}$ values for $^{24}$Mg determined using various probes are consistent with each other within the experimental uncertainties. On the other hand, the hexadecapole deformation value of $^{24}$Mg determined 
using the above probes differs significantly and shows large uncertainties. In the present Letter, results obtained on the ground state $\beta_{2}$ and $\beta_{4}$ values of $^{24}$Mg from QEL scattering at backward angles, are presented. The $\beta_{2}$ value for $^{24}$Mg obtained in the present work shows good agreement with theory and those obtained using inelastic-scattering probes. The ground state hexadecapole deformation value of $^{24}$Mg has been determined with a 95\% confidence limit for the first time. Present results for a light mass nucleus, $^{24}$Mg, along with earlier results in the heavy-mass region of rare earths, establish clearly that QEL scattering is a very useful probe to determine the ground state deformation of exotic nuclei using low intensity radioactive ion beams.
\begin{figure}
\centering\includegraphics[trim= 0.2mm 0.2mm 0.2mm -2mm, clip, height=0.25\textheight]{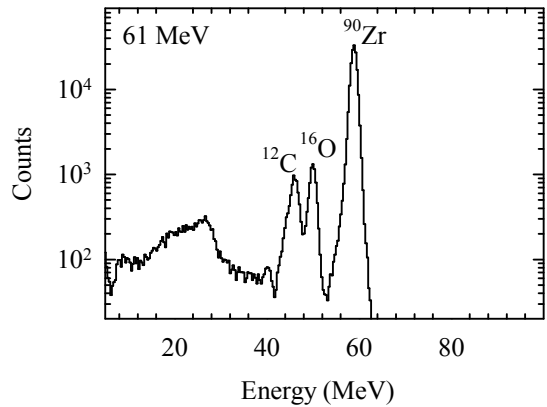}
\caption{\label{24Mg_MonSpect} Rutherford scattering events at monitor angle of 20$^{\circ}$ in $^{24}$Mg + $^{90}$Zr reaction at a beam energy of 61 MeV. The scattering events from $^{12}$C, $^{16}$O, and $^{90}$Zr, are marked.}
\end{figure}

In order to study the deformation effects of $^{24}$Mg, any spherical closed shell target with high charge product of projectile and target ($\mathrm{Z_{P}Z_{T}}$) is an appropriate choice. In the present experiment, considering the available beam energy of $^{24}$Mg, we have opted $^{90}$Zr target with $\mathrm{Z_{P}Z_{T}}$=480 to study the deformation effects in $^{24}$Mg via quasi-elastic measurements in the $^{24}$Mg + $^{90}$Zr reaction.

Quasi-elastic scattering measurements were carried out using $^{16}$O and $^{24}$Mg beams from the FN accelerator facility at the Nuclear Science Laboratory, University of Notre Dame, USA. A
168 $\mu$g/cm$^{2}$ foil of highly enriched ($>$95\%) $^{90}$ZrO$_{2}$ deposited on $^{12}$C (40 $\mu$g/cm$^{2}$) was used as the target. Beam energies were used in the range of  36 to 62 MeV (for $^{16}$O) and 61 to 93 MeV (for $^{24}$Mg) in steps of 1 MeV (for $^{16}$O) and 2 MeV (for $^{24}$Mg). Quasi-elastic scattering events were detected  using three  silicon-surface barrier (SSB) telescopes ($\Delta E-E$) placed at 158.0$^{\circ}$ (17 $\mathrm{\mu m}$, 1 mm), 147.3$^{\circ}$(15 $\mathrm{\mu m}$, 1 mm), and 136.9$^{\circ}$ (23.6 $\mathrm{\mu m}$, 1 mm) with respect to the beam direction. The angular opening of each telescope
was restricted to close to $\pm 1^{\circ }$. Additionally, two SSB detectors each of 1 mm thickness were placed at 126.2$^{\circ }$ and 115.6$^{\circ }$ for the purpose of asserting the quasi-elastic events by kinematic progression. Two more SSB detectors (1 mm) were mounted at 20.0$^{\circ}$ in the reaction plane on either side of the beam direction for the purpose of Rutherford normalization. These two monitor detectors each having a collimator of 2 mm, were placed at a distance of 47.5 cm from the target. At every beam energy change, the transmission of the beam was maximized through a collimator of 5 mm diameter, enabling a halo-free beam. The target $^{90}$Zr possesses certain fraction of $^{16}$O (ZrO$_{2}$) and  $^{12}$C (backing). At forward angles ($\pm$20$^{\circ}$), the Rutherford scattering events were clearly separated for $^{12}$C, $^{16}$O, and $^{90}$Zr as shown in Fig.~\ref{24Mg_MonSpect} for $^{24}$Mg + $^{90}$Zr reaction at a beam energy of 61 MeV.
\begin{figure}[t]
\centering\includegraphics[trim= 0.2mm 0.2mm 0.2mm -2mm, clip, height=0.45\textheight]{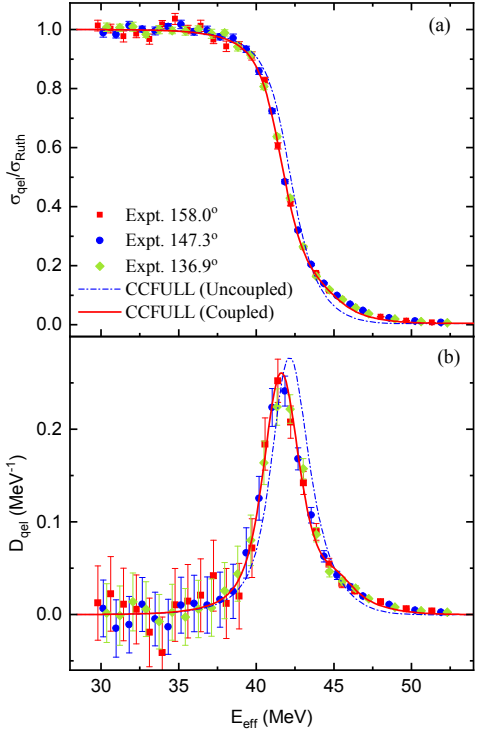}
\caption{\label{ExFunc_BD_16O90Zr} (Color online) Quasi-elastic excitation function (panel (a)) and derived barrier distribution (panel (b)) determined at three backward angles. Dash-dotted and solid lines in both the panels represent the coupled channels calculations using the code CCFULL without including any coupling and with vibrational couplings of $^{90}$Zr (2$^{+}$ and 3$^{-}$ states), respectively (see text).}
\end{figure}

Quasi-elastic events consist of elastic, projectile and target excitation, and to some extent particle transfer events. In the case of $^{16}$O + $^{90}$Zr reaction, the quasi-elastic events were quite evident from the $\Delta E$ versus $E$ plots. However, in the case of $^{24}$Mg + $^{90}$Zr reaction, most of the quasi-elastic events stopped in the $\Delta E$ detectors and only a few events penetrated to the $E$-detector. By putting appropriate two-dimensional gates, it was ensured that quasi-elastic events were free from other light charged particle events. Among quasi-elastic events, the elastic events were dominant. All the SSB detectors were energy calibrated using a $^{229}$Th $\alpha$-source. Successive changes in the kinetic energies of elastic events with varying beam energy were in agreement with two-body kinematics at all angles from 158$^{\circ}$ to 115$^{\circ}$, which further benchmarked the identification of quasi-elastic events. The beam energies were corrected for energy loss in the half-thickness of the target.

Differential cross section for quasi-elastic events at each beam energy was normalized with Rutherford scattering cross section. The center-of-mass energy ($E_\mathrm{c.m.}$) was corrected for centrifugal effects at each angle as follows \cite{Timmers1995, HaginoPRC2004, BKN2007, Piasecki2005}:
\begin{equation}
E_\mathrm{eff} =\frac{2E_\mathrm{c.m.}}{(1+\mathrm{cosec}(\theta_\mathrm{c.m.}/2))}
\end{equation}
\noindent 
where $\theta_\mathrm{c.m.}$ is the center-of-mass angle. The quasi-elastic events in the $^{16}$O + $^{90}$Zr reaction have contributions dominantly from elastic and the target excitations. Owing to large negative $Q$-values, contribution from transfer channels are negligibly small. The quasi-elastic excitation function for the $^{16}$O + $^{90}$Zr reaction is shown in the Fig. \ref{ExFunc_BD_16O90Zr}(a) at the three backward angles. It is seen that the quasi-elastic excitation functions at these backward  angles are overlapping. The quasi-elastic barrier distribution $D_\mathrm{qel}$ ($E_\mathrm{eff}$) from the quasi-elastic excitation function was determined using the relation \cite{Timmers1995}:
\begin{equation}
D_{qel}(E_\mathrm{eff}) = -\frac{d}{dE_\mathrm{eff}}\bigg[ \frac{d\sigma_\mathrm{qel}(E_\mathrm{eff})}{d\sigma_\mathrm{R}(E_\mathrm{eff})}\bigg],
\end{equation}
\noindent 
where $\sigma_\mathrm{qel}$ and $\sigma_\mathrm{R}$ are the differential cross sections for the quasi-elastic and Rutherford scatterings, respectively. A point difference formula is used to 
evaluate the barrier distribution, with the energy step of $\Delta$E=2 MeV in the laboratory frame of reference. Similar to the excitation function, the barrier distribution determined from 
the  excitation functions at three backward angles overlap quite well as shown in the Fig. \ref{ExFunc_BD_16O90Zr}(b) for the $^{16}$O + $^{90}$Zr reaction.

Coupled channels (CC) calculations were carried out for the $^{16}$O + $^{90}$Zr reaction using a modified version of CCFULL code \cite{ccqel} for quasi-elastic scattering. Wood-Saxon type optical model potentials were used for both the real as well as imaginary parts. The optical model parameters (OMPs) for the real potential were grossly estimated from the Broglia-Winther potentials, and those were further refined so that the uncoupled calculation could reproduce the experimental data as best as possible. The OMPs for the real potential used for the $^{16}$O + $^{90}$Zr reaction were as follows: the depth of the potential, $V_{r}$=57.96 MeV, the radius parameter, $R_{r}$=1.2 fm, and  the diffuseness parameter, $a_{r}$=0.585 fm. For the imaginary part of the optical potential, a potential was set to be well confined 
inside the Coulomb barrier in order to simulate a compound nucleus formation. The imaginary potential parameters used in the CC calculations were as follows: the depth of the potential, $V_{I}$=30 MeV, the radius parameter, $R_{I}$=1.0 fm, and  the diffuseness parameter, $a_{I}$=0.09 fm. It is to be noted here that results are not sensitive to the imaginary potential parameters as long as the potential is well confined inside the Coulomb barrier. The radius parameters for the projectile ($R_{P}$) and target ($R_{T}$) were used  to be 1.2 and 1.06 fm, respectively, in the coupled channels Hamiltonian. The Coulomb radius was used to be 1.1 fm.




Using the above potential parameters, calculations were carried out first without including any channel coupling for the $^{16}$O + $^{90}$Zr reaction.
These uncoupled calculations are represented by the dash-dotted lines in Figs. \ref{ExFunc_BD_16O90Zr} (a) and (b). It is clearly seen that uncoupled calculations cannot reproduce the experimental data.
CC calculations were further performed for the $^{16}$O + $^{90}$Zr reaction including the vibrational couplings of the  target, $^{90}$Zr. Excitations in $^{16}$O are not explicitly taken 
into account in the calculations, as they simply renormalize the potential due to the large excitation energies \cite{HaginoPTP2012}.  For the channel couplings to the collective excited
states in the  $^{90}$Zr nucleus, we take into account the vibrational quadrupole (2$^{+}$) state at 2.19 MeV and the octupole (3$^{-}$) state at 2.75 MeV. The deformation
parameters (coupling strengths) associated with the transition of multipolarity $\lambda$ were estimated from measured transition probabilities B(E$\lambda$) \cite{BE2, BE3}. The $\beta_{2}$ and $\beta_{3}$ values used for the 2$^{+}$ and 3$^{-}$ states of $^{90}$Zr were 0.089 and 0.211, respectively \cite{kalkal2005}. Using these $\beta_{\lambda}$ values, CC  calculations were carried out which reproduce the experimental data very well, as shown in Figs. \ref{ExFunc_BD_16O90Zr} (a) and (b). This agreement between  the experimental data and the coupled channels calculations established the reasonableness of the coupling strengths of $^{90}$Zr, which will be used for the $^{24}$Mg + $^{90}$Zr reaction.

\begin{figure}[t]
\centering\includegraphics[trim= 0.2mm 0.2mm 0.2mm -2mm, clip, height=0.45\textheight]{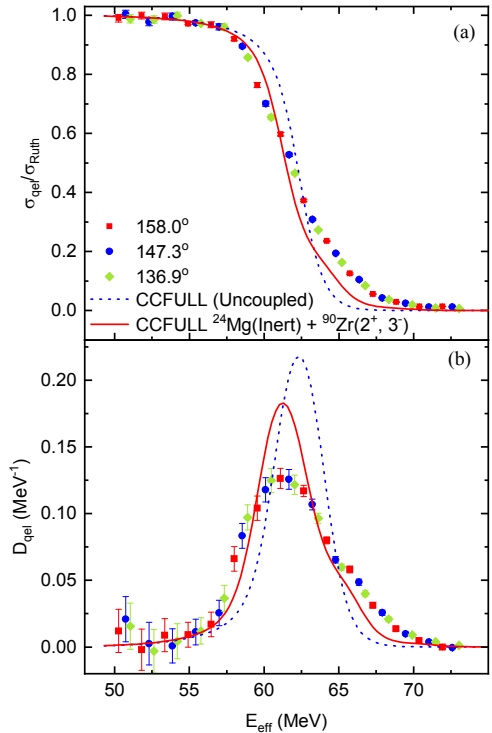}
\caption{\label{24Mg_Expt_BD} (Color online) Quasi-elastic excitation function (panel (a)) and derived barrier distribution (panel (b)) determined at three backward angles for $^{24}$Mg + $^{90}$Zr reaction. The dotted and  solid lines represent CCFULL calculations without including any coupling (uncoupled) and with including vibrational couplings of $^{90}$Zr (2$^{+}$, 3$^{-}$), respectively.}
\end{figure}

The quasi-elastic excitation function and the derived barrier distribution for the $^{24}$Mg + $^{90}$Zr reaction are shown in  Figs. \ref{24Mg_Expt_BD}(a) and (b), respectively. Due to the large negative $Q$-values, contribution from probable transfer channels is negligibly small in $^{24}$Mg + $^{90}$Zr reaction as revealed in Ref. \cite{Wojcik2018} . The shape of the quasi-elastic excitation function for $^{24}$Mg + $^{90}$Zr reaction does not show any discernible difference when compared with that of $^{16}$O + $^{90}$Zr reaction as shown in the Fig. \ref{ExFunc_BD_16O90Zr}(a). However, as discussed earlier, the derived barrier distribution shows more fingerprints of the couplings of the relative motion with the internal degrees of freedom and it is quite evident while comparing the experimental barrier distribution for $^{24}$Mg + $^{90}$Zr reaction (Fig. \ref{24Mg_Expt_BD}(b)) with that of  $^{16}$O + $^{90}$Zr reaction (Fig. \ref{ExFunc_BD_16O90Zr}(b)). It is noted that the barrier distribution for $^{24}$Mg + $^{90}$Zr reaction does not reveal any sharp structure, but is significantly broader than that of the $^{16}$O + $^{90}$Zr reaction, indicating stronger ground state deformation effects of $^{24}$Mg.

The OMPs for the real potential used for the $^{24}$Mg + $^{90}$Zr reaction were as follows: the depth of the potential, $V_{r}$= 160.0 MeV, the radius parameter, $R_{r}$=1.1 fm, and  the diffuseness parameter, $a_{r}$=0.620 fm. For the imaginary part of the optical potential, the same potential parameters were used as those used for the $^{16}$O + $^{90}$Zr reaction except for the diffuseness parameter, $a_{I}$=0.1 fm. 
The radius parameters used for the projectile ($R_{P}$) and target ($R_{T}$) in the coupled channel Hamiltonian were 1.2 and 1.06 fm, respectively. The Coulomb radius used was 1.1 fm.


Using the above potential parameters, at first, CCFULL calculations for  $^{24}$Mg + $^{90}$Zr reaction were carried out without including any channel coupling.
These uncoupled calculations are represented by the dotted lines in Figs. \ref{24Mg_Expt_BD} (a) and (b). It is clearly seen that uncoupled calculations cannot reproduce the experimental data.
CC calculations were further performed by including the vibrational couplings of the  target, $^{90}$Zr, while the projectile, $^{24}$Mg was treated as an inert nucleus. For the channel couplings to the collective excited states in the  $^{90}$Zr nucleus, we took into account the vibrational quadrupole (2$^{+}$) state at 2.19 MeV and the octupole (3$^{-}$) state at 2.75 MeV as determined from the $^{16}$O + $^{90}$Zr reaction. These calculations are shown by solid lines in Figs. \ref{24Mg_Expt_BD} (a) and (b). It is clearly seen that considering $^{24}$Mg as an inert nucleus, CCFULL calculations deviate significantly from the experimental data, raising the urge to include the rotational couplings of $^{24}$Mg within the CCFULL framework.

Rotational degrees of freedom were included in the CCFULL calculations in order to reproduce the quasi-elastic excitation function and the barrier distribution for $^{24}$Mg + $^{90}$Zr reaction. The rigid rotor model was used for this purpose. At first only quadruple deformation ($\beta_{2}$) was considered and calculations were performed with various values of $\beta_{2}$ in the range of 0.2 to 0.6, keeping vibrational couplings of $^{90}$Zr as determined earlier. The first three rotational states of $^{24}$Mg (0$^{+}$, 2$^{+}$, and 4$^{+}$) were included in the CCFULL calculations. The coupling to the 6$^{+}$ state has been confirmed to give a negligible contribution. The Coulomb ($\beta^C_2$) and nuclear ($\beta^N_2$) parts of the quadrupole deformation were kept at the same values. It is observed that ground state quadrupole deformation alone cannot reproduce the experimental data in the full energy range.

As discussed earlier,  $^{24}$Mg shows the signature of non-zero hexadecapole  deformation in its ground state \cite{Horikawa1971, Akira1972, swiniarski1969, Leo1979, Peterson1978}. It has been revealed through several experimental investigations using electron-, proton, $^{3}$He, $\alpha$ scattering, and is supported by microscopic theories \cite{MollerNix1995, Yoshida_pvt}. However, previously determined ground state hexadecapole deformation parameter ($\beta_{4}$) of $^{24}$Mg using various probes varies quite dramatically and possess large uncertainties. In order to reproduce the present experimental data, hexadecapole deformation has also been included along with the quadrupole deformation. CCFULL calculations were carried out in the two dimensional space of $\beta_{2}$  and $\beta_{4}$ of  $^{24}$Mg, considering the first three rotational states (0$^{+}$, 2$^{+}$, and 4$^{+}$). The Coulomb and nuclear parts for both quadrupole and hexadecapole deformations were kept at same values. The $\beta_{2}$ values were varied in the range of 0.2 to 0.6 in a step of 0.01, and for each value of the $\beta_{2}$, the $\beta_{4}$ was varied in the range of -0.20 to +0.20 with a step size of 0.01. Vibrational coupling strengths of $^{90}$Zr were used as determined earlier from $^{16}$O +$^{90}$Zr scattering.

\begin{figure}
\centering\includegraphics[trim= 0.2mm 0.2mm 0.2mm -2mm, clip, height=0.25\textheight]{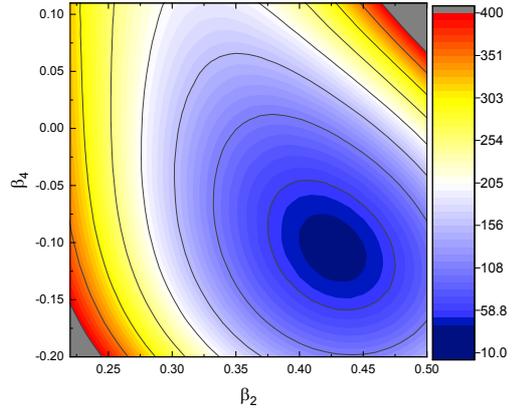}
\caption{\label{Chisq} (Color online) $\chi^{2}$ distribution in the two dimensional space of $\beta_{4}$ versus $\beta_{2}$ of $^{24}$Mg, determined by comparing experimental barrier distribution with CCFULL calculations (see text).}
\end{figure}

$\chi^{2}$ was calculated between the experimental data (for the barrier distribution) and CCFULL calculation for each combination of $\beta_{2}$ and $\beta_{4}$ using the following equation;
\begin{equation}
\chi^{2}(\beta_{2},\beta_{4})=\sum_{i=1}^{N}\frac{[Y_{i}-f(\beta_{2},\beta_{4})]^2}{\sigma_{i}^{2}}
\label{Chi}
\end{equation}
\noindent
where $Y_{i}$ represents the experimental value of the barrier distribution at the $i^{th}$ energy point, $\sigma_{i}$ is the uncertainty in the data, and $f(\beta_{2},\beta_{4})$ represents the corresponding CCFULL calculation for a particular combination of $\beta_{2}$ and $\beta_{4}$. In  Eq. \ref{Chi}, the summation runs over all the data points ($N$) in the effective energy $E_\mathrm{eff}$  range from 50 to 73 MeV. The $\chi^{2}$-distribution thus obtained in the two-dimensional space of $\beta_{4}$ versus $\beta_{2}$ is shown in Fig. \ref{Chisq}. It is seen that for a very small region in the two-dimensional space of $\beta_{2}$ and $\beta_{4}$ (see Fig. \ref{Chisq}), $\chi^{2}$ is minimized.

In order to get the quantitative values of $\beta_{2}$ and $\beta_{4}$  and their associated uncertainties, a Bayesian analysis with a Markov-Chain Monte Carlo (MCMC) framework was carried out. The aforementioned $\chi^2$ distribution simultaneously constrains the likelihood function, which is defined as
\begin{equation}
  P(\vec{Y} | \beta_2, \beta_4) = \exp \left ( - \chi^2 /2 \right).
\end{equation}
The likelihood function is a conditional probability density of a dataset, $\vec{Y}$, given some values for the model parameters $\beta_2,\beta_4$. In turn, the inverse conditional probability, $P(\beta_2, \beta_4 | \vec{Y})$ yields information on the distribution of $\beta_2$ and $\beta_4$ given a set of data. The connection between these two probability distributions is encapsulated within Bayes' Theorem:
\begin{equation}
  P(\beta_2, \beta_4 | \vec{Y})  = \frac{ P(\vec{Y} | \beta_2, \beta_4 ) P(\beta_2,\beta_4) } {P(\vec{Y}) }.
\label{Bayes}
\end{equation}

\noindent
In Eq. \ref{Bayes}, $P(\vec{Y})$ and $P(\beta_2,\beta_4)$ are, respectively, the so-called prior distributions of $\vec{Y}$ and $(\beta_2,\beta_4)$ which were merely taken to be uniform distributions over the parameter space. However, during the MCMC simulation, as the values of $\beta_2$ and $\beta_4$ change, the value of $P(\vec{Y})$ is constant. At each step of the simulation, Eq. \ref{Bayes} is evaluated for each value of $\beta_2$ and $\beta_4$, and compared with the value of Eq. \ref{Bayes} of the previous step. 

It is Eq. \ref{Bayes} which allows for one to infer the probability distributions of the parameters $\beta_2$ and $\beta_4$ from experimental data. 
The Python implementation of the affine-invariant algorithm of Goodman and Weare was used \cite{goodman_weare, foreman_mackey}. In this algorithm, $1000$ ``walkers'' were randomly initialized in locations in the two-dimensional parameter space of ($\beta_2$, $\beta_4$). In parallel, these walkers took Markovian steps which were accepted subject to the value of Eq. \ref{Bayes} and the MCMC criteria \cite{goodman_weare}.

\begin{figure}
  \centering\includegraphics[trim= 0.2mm 0.2mm 0.2mm -2mm, clip, height=0.35\textheight, angle=270]{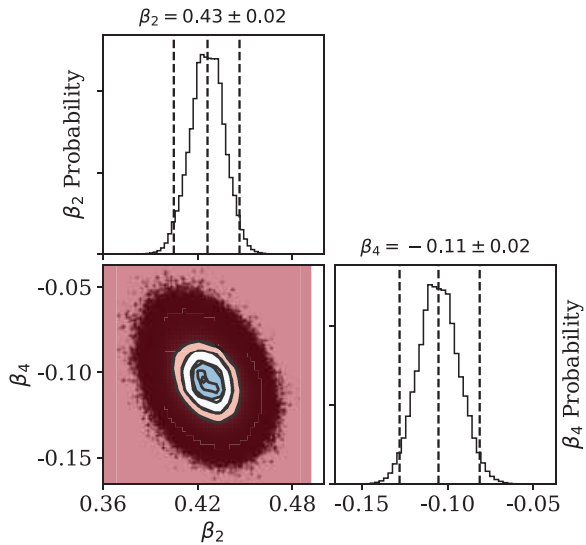}
  \caption{Multidimensional probability distributions for $\beta_2$ and $\beta_4$ resulting from the MCMC simulation from the experimental data (see text). Plus- and minus-uncertainties are shown and constitute a $95\%$ confidence interval in the data.}
  \label{probability_dists}
\end{figure}

These features of the Bayesian analysis yield, after convergence is reached, histograms of the walker positions which converge to the posterior distribution of the parameter space. These resulting probability distributions are shown in Fig. \ref{probability_dists}. The $\beta_2$ and $\beta_4$ are moderately anticorrelated with a correlation of $\sim - 0.298$, which is shown graphically within the two-dimensional probability distribution within Fig. \ref{probability_dists}. Examination of the projections of the probability density onto the parameter axes yields extracted values of $\beta_2 = +0.43 \pm 0.02$ and $\beta_4 = - 0.11 \pm 0.02$, with approximately symmetric distributions centered at the medians. The uncertainties  constitute a $95\%$ confidence interval in the data.

\begin{figure}
\centering\includegraphics[trim= 0.2mm 0.2mm 0.2mm -2mm, clip, height=0.28\textheight]{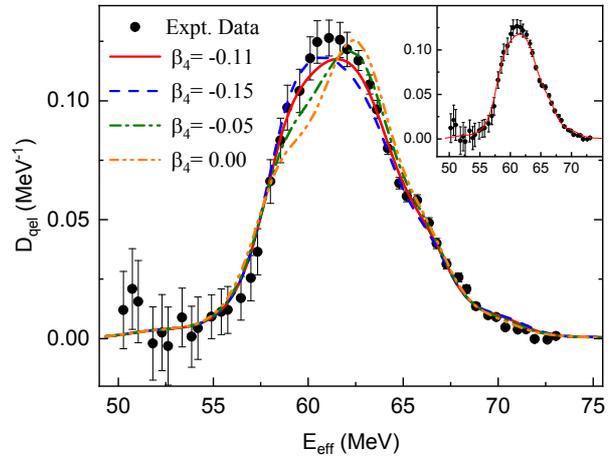}
\caption{\label{BD_FinalB2B4} (Color online) The quasi-elastic barrier distribution for $^{24}$Mg + $^{90}$Zr reaction. Different lines represent CCFULL calculations with fixed quadrupole ($\beta_{2}$=+0.43) and varying hexadecapole deformation parameters ($\beta_{4}$) of $^{24}$Mg. Solid (red), dashed (blue), dash-dotted (green), and dash-dot-dotted (orange) lines correspond to $\beta_{4}$=-0.11, -0.15, -.05 and 0.00, respectively. The inset shows barrier distribution data and calculations only for $\beta_{2}$=+0.43 and $\beta_{4}$=-0.11. }
\end{figure}

The experimental data for the barrier distribution were compared with CCFULL calculations as shown in Fig. \ref{BD_FinalB2B4} using the $\beta_{2}$ and $\beta_{4}$ values of $^{24}$Mg  as determined from the above Bayesian analysis. CCFULL calculations with various $\beta_4$ values and fixed $\beta_2$=+0.43 are also shown in the Fig. \ref{BD_FinalB2B4}. The barrier distribution shows good sensitivity with  $\beta_4$ as depicted in the Fig. \ref{BD_FinalB2B4}. The inset of Fig. \ref{BD_FinalB2B4} shows barrier distribution data and calculations only for $\beta_{2}$=+0.43 and $\beta_{4}$=-0.11. One can see that within the experimental uncertainties,  CCFULL calculations with $\beta_2$=+0.43 and $\beta_4$=-0.11 reproduce the barrier distribution very well. 

The $\beta_2$ and $\beta_4$ values of $^{24}$Mg determined in the present work using quasi-elastic scattering have been compared in Table \ref{table3} with those reported earlier in the literature. It is seen from Table \ref{table3} that except neutron-scattering, the $\beta_2$ value determined in the present work shows a good overlap with those determined using different inelastic scattering probes. This value also shows a close proximity with the theoretical values provided in the Table \ref{table3}. Using inelastic scattering probes, the hexadecapole deformation parameter $\beta_4$,  either had no quoted error or the uncertainties were quite large.  Moreover, previously determined $\beta_4$ values of $^{24}$Mg vary quite dramatically as also shown in the Table \ref{table3}. It is the first time that $\beta_4$ of $^{24}$Mg has been determined with a 95\% confidence limit to be -0.11$\pm$0.02. The present results along with earlier work \cite{Jia2014} in the heavy mass region, clearly establish that quasi-elastic scattering is a sensitive probe to determine the ground state deformation parameters.


In summary, quasi-elastic measurements have been performed for the $^{16}$O + $^{90}$Zr and $^{24}$Mg + $^{90}$Zr reactions at different laboratory angles. Quasi-elastic excitation function and the derived barrier distributions therefrom were compared with Coupled Channels (CC) calculations using the code CCFULL. Vibrational channel coupling strengths of $^{90}$Zr were obtained from $^{16}$O + $^{90}$Zr reactions which were found to be consistent with literature data. Rotational channel couplings of $^{24}$Mg were required to reproduce the experimental data for the $^{24}$Mg + $^{90}$Zr reaction by the CC calculations. The best choice of ground state quadrupole ($\beta_2$) and hexadecapole ($\beta_4$) deformation parameters for $^{24}$Mg  was searched for using Bayesian analysis. The $\beta_2$ value obtained  for $^{24}$Mg shows good consistency with previously reported data and microscopic theories. Data for $^{24}$Mg + $^{90}$Zr reaction shows very good sensitivity to hexadecapole deformation of $^{24}$Mg, and a precise experimental value (with 95\% confidence limit) has been obtained for the first time. 

We point out that a quasi-elastic barrier distribution is
especially useful with radioactive beams, with which
high precision measurements for fusion cross sections would
be difficult in order to extract a fusion barrier distribution.
This is also the case for fusion reactions relevant to superheavy
elements \cite{Ntshangase2007,Tanaka2018,Tanaka2020}.
The present results shown in this Letter clearly demonstrate
that quasi-elastic scattering could be a potential probe
to determine the ground state deformation of the exotic
nuclei using low intensity radioactive ion beams.

\begin{table}
\caption{\label{table3}Quadrupole and hexadecapole deformation of $^{24}$Mg using different experimental probes and theoretical calculations.}
\begin{tabular} {lll}
\hline
Probe & $\beta_2$  & $\beta_4$ \\
\hline
\\
Present Work                        & +0.43 $\pm$0.02 & - 0.11 $\pm$ 0.02\\

(e, e$'$ ) \cite{Horikawa1971}  & +0.45                      & -0.06\\
(e, e$'$ )  \cite{Akira1972}    & +0.47 $\pm$ 0.03           & -0.03\\

(p, p$'$ )              \cite{swiniarski1969} & +0.47           & - 0.05 $\pm$ 0.08\\
(p, p$'$ )               \cite{Leo1979} &        +0.486 $\pm$0.008           & +0.05 $\pm$ 0.04\\

(n, n$'$ )              \cite{Haouat1984}  & +0.50 $\pm$0.02         &   0.00 $\pm$ 0.01 \\

(d, d$'$ )               \cite{TJIN1968}      & +0.42           &   \\
(d, d$'$ )               \cite{KISS1969}      & +0.40           &   \\

($^{3}$He, $^{3}$He$'$ )\cite{Peterson1978}   & +0.42 $\pm$ .04           & -0.02$^{+.01}_{-.02}$   \\

($\alpha$, $\alpha '$ )  \cite{Rebel1972}     & +0.39 $\pm$ .01           &  -0.015 $\pm$  0.015  \\
($\alpha$, $\alpha '$ )  \cite{Harakeh1979}   & +0.355                     & -0.03\\


\\
FRDM\footnotemark[1]                   \cite{MollerNix1995}   & $\beta^N_2$=+0.374           & -0.053   \\

Skyrme HFB\footnotemark[1]                                        \cite{Yoshida2010}           & $\beta^N_2$=+0.40     &    \\
                                                                                     & $\beta^C_2$=+0.41     &    \\
\hline

\end{tabular}
\footnotemark[1]{Theory}  
  
\end{table}

YKG is thankful to Dr. R. K. Choudhury for discussion at various stages of this work. This work has been supported in part by the US National Science Foundation (Grant Nos. PHY1419765 and PHY1762495).


\end{document}